# Spin Orbit Torque Based Electronic Neuron

Abhronil Sengupta[a,*], Sri Harsha Choday[*], Yusung Kim, and Kaushik Roy

School of Electrical & Computer Engineering, Purdue University, West Lafayette, IN, 47906, USA

**Abstract:** A device based on current-induced spin-orbit torque (SOT) that functions as an electronic neuron is proposed in this work. The SOT device implements an artificial neuron's thresholding (transfer) function. In the first step of a two-step switching scheme, a charge current places the magnetization of a nano-magnet along the hard-axis i.e. an unstable point for the magnet. In the second step, the SOT device (neuron) receives a current (from the synapses) which moves the magnetization from the unstable point to one of the two stable states. The polarity of the synaptic current encodes the excitatory and inhibitory nature of the neuron input, and determines the final orientation of the magnetization. A resistive crossbar array, functioning as synapses, generates a bipolar current that is a weighted sum of the inputs. The simulation of a two layer feed-forward Artificial Neural Network (ANN) based on the SOT electronic neuron shows that it consumes ~3X lower power than a 45nm digital CMOS implementation, while reaching ~80% accuracy in the classification of one hundred images of handwritten digits from the MNIST dataset.

Artificial neural networks (ANNs) attempt to replicate the remarkable efficiency of the biological brain for performing cognitive tasks such as learning, pattern recognition and classification [1-2]. At the heart of any ANN is an artificial neuron whose transfer function mimics that of a biological neuron. One of the most widely used models of an artificial neuron with an output ($y$) and a transfer function ($f$) can be written as $y = f(\sum_i w_i . x_i + b)$ where, $x_i$ is an input to the neuron, $w_i$ is its corresponding synaptic weight, and $b$ is a constant bias term. Thus the two main computational units of the artificial neuron are weighted summation of inputs followed by a thresholding operation. Traditionally, ANNs have been implemented in software running on a Von-Neumann type general-purpose computer [3]. The implementation of large scale ANNs on general purpose computers requires significant computational capability and consumes energy that is orders of magnitude larger than its biological counterpart [3]. The recent developments in the field of neuromorphic computation attempt to bridge this gap by emulating artificial neurons using custom analog/digital CMOS circuits [4-5]. However, the emulation of artificial neurons using CMOS circuits remains highly inefficient in terms of energy consumption and silicon area [6]. The inefficiency in CMOS based ANNs arises due to the significant mismatch between the functionality of a biological neuron and the CMOS devices which are better suited for Boolean logic.

---

a) Corresponding Author email: asengup@purdue.edu

*Authors contributed equally to this work.

Recent discoveries in Spintronics and Resistive memories have produced devices that can enable a direct mapping of neuron functionality to the electronic implementation, thus creating possibilities for energy-efficient ANN hardware. For example, several nano-magnets can inject weighted spin currents into a non-magnetic channel whose spin potential depends on the sum of all the injected spin currents. Such functionality directly corresponds to the synaptic operations in the biological domain. Moreover, when the net spin potential in the channel exceeds a certain threshold, it can switch an output magnet. This represents the thresholding operation in the biological domain. Recently, Sharad *et al.* proposed neuromorphic hardware utilizing Lateral Spin Valves (LSV) and Domain Wall Magnets (DWM) [6-7] which showed significant energy improvements compared to the CMOS implementation of an ANN. In addition to spin domain neural hardware, memristive [8-10] and oxide-based resistive switching devices [11] have also shown promising results for the implementation of synaptic functionality.

Recently, the use of current-induced spin-orbit torque (SOT) has attracted significant attention due to its capability in achieving a spin injection efficiency >1 [14-21]. This spin-orbit torque occurs in a ferromagnet/heavy metal (FM/HM) bilayer structure in which the flow of charge current through a heavy metal results in an efficient control of a thin nano-magnet placed on top. Although the origin of the spin-orbit torque is still being debated, the potential benefit of using SOT has been repeatedly confirmed by several experiments [17,18,19,24].

In this work, we propose a device based on current-induced SOT that functions as the thresholding unit of an electronic neuron. In the first step of a two-step switching scheme, a charge current through HM places the magnetization of a nano-magnet along the hard-axis i.e. an unstable point for the magnet. In the second step, the device receives a current (from the electronic synapse) which moves the magnetization from the unstable point to one of the two stable states. The polarity of the net synaptic current determines the final orientation of the magnetization. A resistive crossbar array, as discussed later in the paper, functions as the synapse generating a bipolar current that is a weighted sum of the inputs.

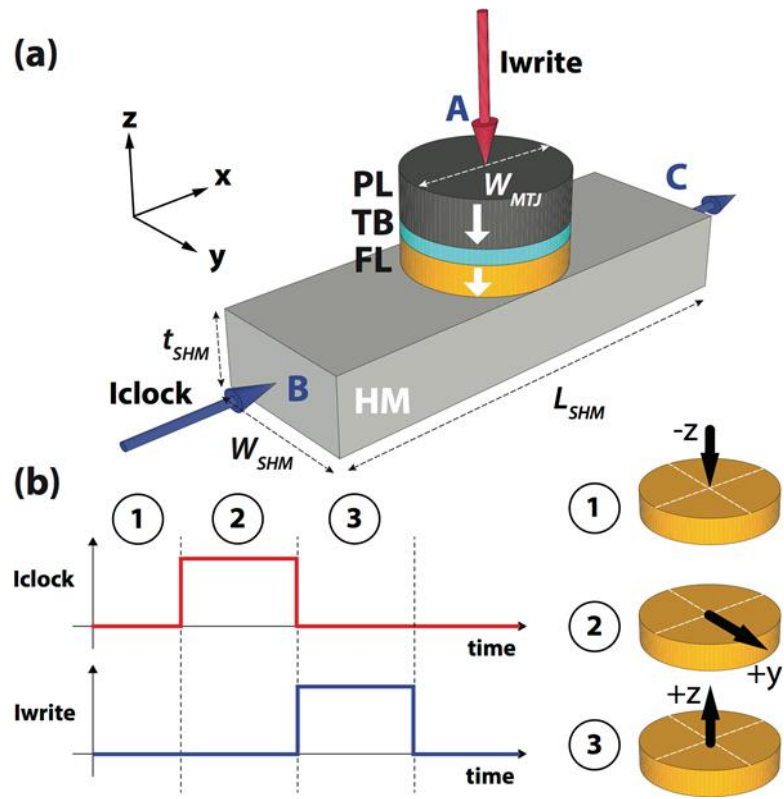

**Fig. 1: (a) Thresholding device for spin-neuron (b) two-step switching scheme (c) magnetization directions of the nanomagnet.**

The proposed thresholding device is shown in Fig. 1(a). It consists of a heavy metal (HM) with high spin-orbit coupling, and a Perpendicular Magnetic Anisotropy (PMA) free layer (FL) in contact with the top surface of the HM. The energy landscape of FL with uniaxial anisotropy (which could originate from shape, interface, or bulk magneto-crystalline anisotropy) is shown in Fig. 2 where the two energy minima points (stable magnetization points) are separated by an anisotropy barrier. In addition, the FL is part of a magnetic tunnel junction (MTJ) consisting of an oxide tunnel barrier (TB) and a PMA pinned layer (PL).

In the HM layer, when spin Hall effect (SHE) [14,18,19] is the dominant underlying physical mechanism in play, a flow of charge current through the HM generates pure spin current in the direction transverse to the charge current due to preferential scattering of different spins to different directions. This pure spin current is then used to control the FL on top, via spin-transfer torque effect [12,13].

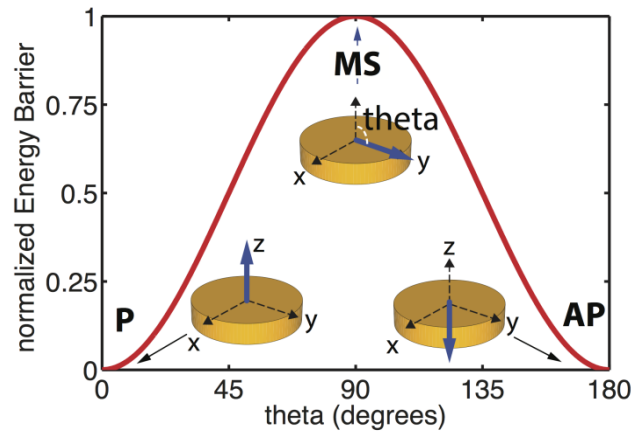

**Fig. 2: Normalized energy landscape of a nanomagnet with a uniaxial anisotropy in out-of-plane direction.**

The proposed thresholding device utilizes a two-step switching scheme in order to a) minimize the required current from the electronic synapse, and b) utilize SHE for an energy efficient thresholding operation. Two-step switching schemes have been utilized previously in magnetic quantum-dot cellular automata (MQCA) [22], All-Spin Logic (ASL) [26], and SHE-assisted-memory bit-cell [29]. As illustrated in Fig. 1(b), for the first step, a charge current ($I_{Clock}$) is supplied through the HM (between terminals B and C) which generates a torque to align the FL magnetization in ±y direction. In other words, $I_{Clock}$ aligns the FL magnetization along the hard-axis of the magnet i.e. the unstable point in the energy landscape (labeled as MS in Fig. 2). Let us define this switching stage as "hard-axis switching". Subsequently in the second step, the electronic synapses drive a charge current ($I_{write}$) between terminals A and C, as illustrated in Fig. 1(b). The net synaptic current ($I_{write}$) flowing through the MTJ exerts a torque on the magnetization which will align the magnet to either one of the easy axis direction along (±z). This step is referred to as "easy-axis switching". The direction of torque generated by $I_{write}$ depends on the polarity of the net synaptic current. If the synaptic current is a positive value, the sign of torque is such that the FL's magnetization becomes anti-parallel (AP) to that of the PL. On the other hand, a negative synaptic current places the FL's magnetization parallel (P) to that of PL. The P and AP states of the MTJ correspond to the low and high (binary) outputs of the neuron. The proposed thresholding device is functionally similar to a biological neuron 'firing' a pulse when the synaptic signal exceeds a certain threshold.

The direction and the magnitude of spin current and its spin polarization in SHE can be determined from the relationship, $\boldsymbol{J}_S = \theta_{SH}(\boldsymbol{\sigma} \times \boldsymbol{J}_q)$ [14,18,19] where $\boldsymbol{J}_S$ and $\boldsymbol{J}_q$ are the transverse spin current and charge current, respectively, $\theta_{SH}$ is a material-dependent spin Hall angle, and $\boldsymbol{\sigma}$ is the polarization of the spin current. Magnetization dynamics of the FL are obtained by solving Landau-Lifshitz-Gilbert equation with additional term to account for the torque due to transverse spin current [27]

$$\frac{d\hat{\boldsymbol{m}}}{dt} = -\gamma(\hat{\boldsymbol{m}} \times \boldsymbol{H}_{eff}) + \alpha\left(\hat{\boldsymbol{m}} \times \frac{d\hat{\boldsymbol{m}}}{dt}\right) + \frac{1}{qN_S}(\hat{\boldsymbol{m}} \times \boldsymbol{I}_s \times \hat{\boldsymbol{m}}) \qquad (1)$$

where $\hat{\boldsymbol{m}}$ is the unit vector of free layer magnetization, $\gamma = 2\mu_B\mu_0/\hbar$ is the gyromagnetic ratio for electron [rad.m/(A.s)], $\boldsymbol{H}_{eff}$ is the effective magnetic field [A/m], and $\boldsymbol{I}_S = \theta_{SH}(A_{MTJ}/A_{HM})I_q\boldsymbol{\sigma}$ is the spin current injected into the free-layer [A]. $N_S$ is the number of spins in the free layer defined as $M_S V/\mu_B$ where $M_S$ is saturation magnetization [A/m], $V$ the volume of the free layer (m$^3$), and $\mu_B$ the Bohr magneton (A.m$^2$). The effective field $\boldsymbol{H}_{eff}$ includes shape anisotropy field $\boldsymbol{H}_{shape} = -(N_{XX}, N_{YY}, N_{ZZ})M_S$ with the demagnetization factors $N_{XX}, N_{YY}, N_{ZZ}$ for elliptical disks calculated using [28], magnetocrystalline anisotropy $H_{Ku2}$ perpendicular to the FL plane direction, external magnetic field, $\boldsymbol{H}_a$, and thermal fluctuation field $\boldsymbol{H}_{thermal}$ given by [26]

$$\boldsymbol{H}_{i,thermal}(t) = \sqrt{\frac{\alpha}{1+\alpha^2}\frac{2k_B T}{\gamma\mu_0 M_S V \delta_t}} G_{0,1}$$ where $G_{0,1}$ is a Gaussian distribution with zero mean and unit standard deviation, $k_B$ is the Boltzmann constant, $T$ is the temperature, $\delta_t$ is the simulation time-step, chosen as *0.1ps* in this work.

To determine the appropriate magnitude of clock and write currents for the proposed device, the switching phase diagram for a range of clock and write currents is constructed as shown in Fig.3. For each set of clock and write currents, ~100,000 stochastic LLG simulations were carried out to obtain the statistics of switching. For simplicity, the rise and fall times of the pulses were set to zero and the pulse width for clock and write currents are set to 2ns and 1ns respectively. As it can be observed from the figure, when clock current is large enough, the amount of write current needed to achieve successful switching is on the order of few µA, just enough to overcome thermal fluctuations and tilt the magnet in the desired direction. Thus the proposed device facilitates fast and energy-efficient threshold operation by utilizing Spin-Hall effect for "hard–axis switching" and minimal synaptic current for deterministic "easy-axis switching".

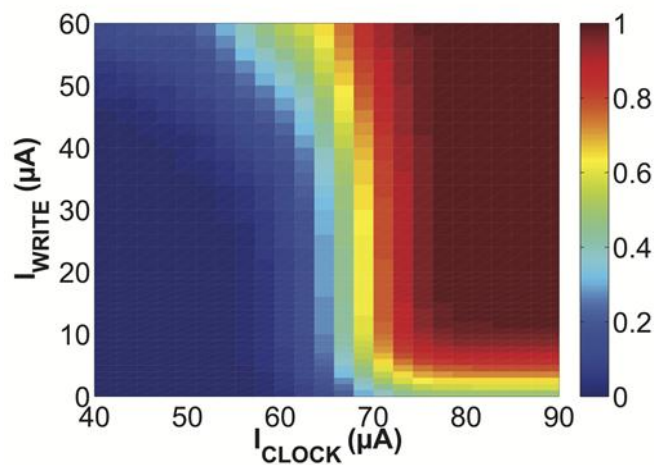

**Fig. 3: Switching phase diagram showing probability of switching for a range of clock and write currents. The FL has the dimension of 40nm x 40nm x 1.5nm.**

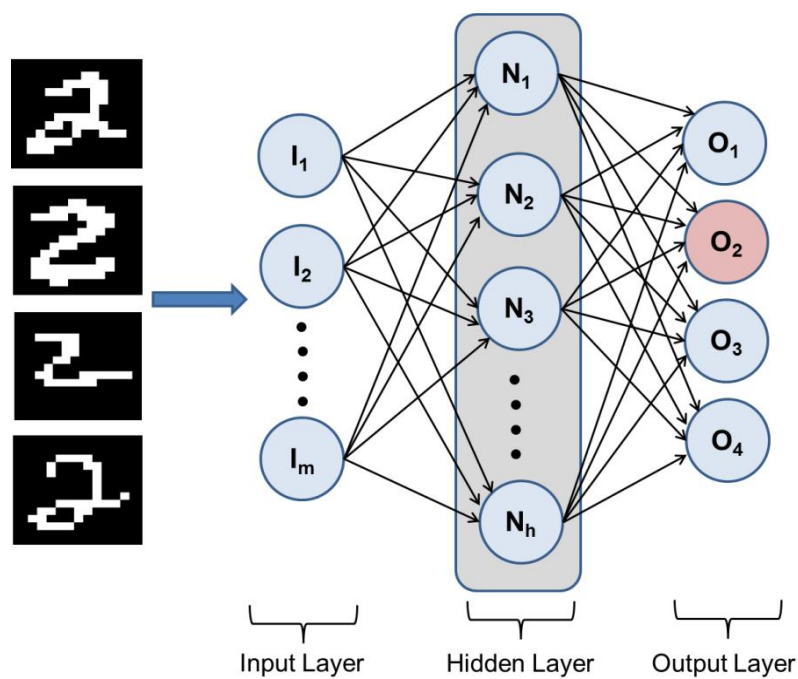

**Fig. 4: Neural network configuration used for digit recognition. The second output neuron gets activated for different variations of digit "2".**

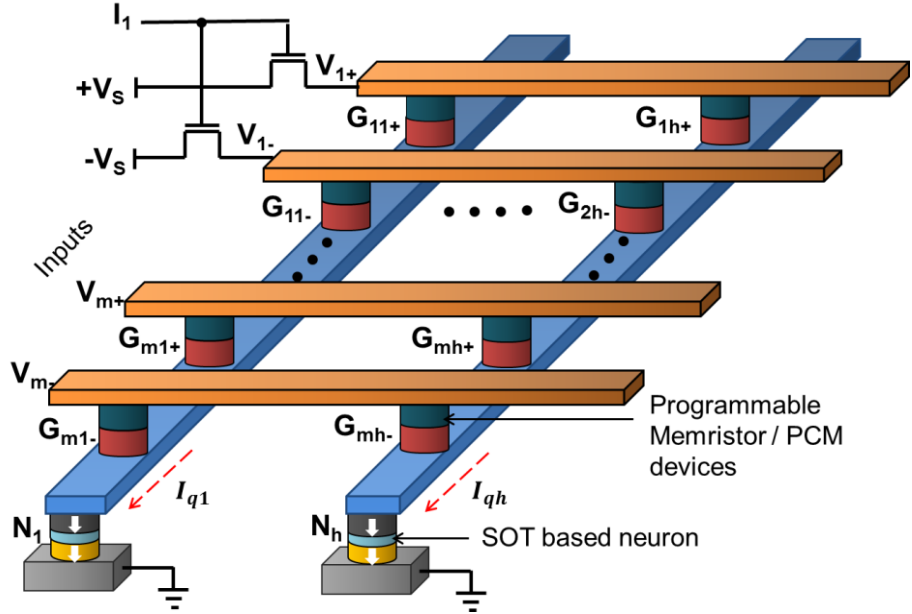

**Fig. 5: Resistive crossbar network for computation of weighted summation of neuron inputs.**

The configuration of a neural network with *m* number of inputs and *h* number of hidden layer neurons is shown in Fig. 4. Let us consider the hardware mapping for the hidden layer of the neural network (shown in Fig. 5). The layer can be represented by a resistive crossbar network (RCN) where each row provides the corresponding input to all the neurons in that layer while each column provides the net synaptic current to the spin neurons. In order to implement bipolar weights, two rows are used for each input. Each input drives the gates of the two pass transistors, one of whose terminals are connected to the RCN while the other terminal is connected to a positive ($+V_s$) and negative ($-V_s$) supply respectively. For efficient operation of the RCN, the input switches should have a very low ON resistance to minimize the voltage drop across them. If the weight for a particular input is positive, then the conductance corresponding to $+V_s$ is programmed to the corresponding weight, while the conductance corresponding to $-V_s$ is programmed to a very high OFF resistive state and vice versa. Each column of the RCN is connected to the spin neuron which receives the resultant synaptic current from the RCN. Considering the conductance in the path of the net synaptic current while flowing through the spin-neuron to be $G_s$, the charge current received by the *j*-th neuron is given by, $I_{qj} = G_s \cdot \sum_{i=1}^{m}(G_{ij+} \cdot V_{i+} + G_{ij-} \cdot V_{i-})/(G_s + \sum_{i=1}^{m}(G_{ij+} + G_{ij-})) \propto \sum_{i=1}^{m}(G_{ij+} \cdot V_{i+} + G_{ij-} \cdot V_{i-})$. Therefore, the charge current $I_{qj}$ is proportional to the weighted summation of the inputs ($V_i$) and the synaptic weights ($G_{ij}$). The sign of the charge current determines the direction of the resultant spin current and hence the final state of the nano-magnet in the SOT-based neuron.

Interlayer communication is performed using the read circuit shown in Fig. 6. A reference MTJ in the AP state is utilized to form a resistive divider network which drives the inverter. For a positive charge current input to the neuron, the MTJ in SOT-based neuron is switched to the AP state. Thus the output voltage after the inverter stage switches to $V_{DD}$ which drives the gates of the two input pass transistors of the RCN in the succeeding stage.

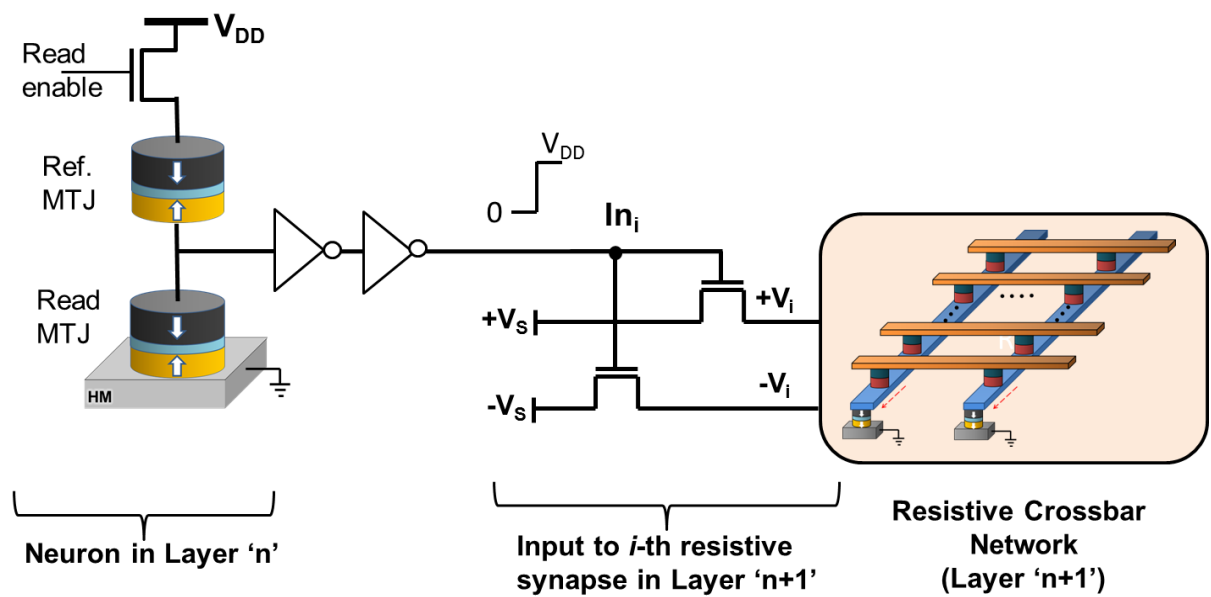

**Fig. 6: Read circuit for neuron state: Interlayer communication.**

For this work, the neural network was designed to recognize the first 4 digits from the MNIST dataset [31]. The images were downscaled to size 8 X 8 and 100 images were utilized for evaluating the performance of the network. The network consisted of 25 hidden layer neurons and 4 output layer neurons. The weights and biases obtained from offline training of the network were mapped to conductance values of a resistive crossbar network. Input currents obtained from SPICE simulations of the RCN were then used to solve stochastic magnetization dynamics for the SOT based neuron. For the first stage of the switching process, a charge current of ~85 µA (from Fig. 3) was used to orient the nano-magnet in the hard-axis position within a duration of 2ns, resulting in a power consumption of ~7.22µW per neuron. The fast and energy efficient "hard-axis switching" is mainly attributed to a spin injection efficiency of 4.71 resulting from SOT. In the next step, the net synaptic charge current from the RCN drives the magnet to one of its stable magnetization states. The operating supply voltages of the RCN were limited by the minimum current required to deterministically switch the spin neuron in the appropriate direction (Fig. 3). For this work, the supply voltages were maintained at ±1 V and the network accuracy was determined by running 100 stochastic simulations of the network for each input image from the total set of 100 images of the dataset. This was performed utilizing the switching probability curve obtained from Fig.3 to capture the effect of the magnet's non-deterministic switching on the accuracy of the network. The accuracy of the network over the set of 10,000 simulation runs was ~80%. Simulation of the read circuit for the neuron state was performed using the NEGF based transport simulation framework proposed in [30]. The average power consumption per neuron obtained from SPICE simulations of the neural network was ~351.94µW.

TABLE I. Simulation Parameters for SOT-based neuron

| Parameters | Value |
|---|---|
| Free layer volume | $\frac{\Pi}{4}$ X 40 x 40 x 1.5 nm$^3$ |
| Saturation Magnetization | 1000 kA/m |
| Spin Hall angle | 0.3 |
| Spin Hall metal dimension | 40 x 40 x 2 nm$^3$ |
| Spin Hall metal resistivity | 200 µΩ-cm |
| Gilbert Damping Factor | 0.0122 |
| Energy Barrier | 31.44 KT |
| MgO Thickness | 1.0 nm |
| Programming range of RCN | 8-160 KΩ |
| No. of programming levels in RCN | 32 |
| RCN operating voltage | 1 V |

In order to perform an iso-throughput comparison with digital CMOS technology, a neural network hardware was synthesized using a standard cell library in 45nm commercial CMOS technology. A 5-bit precision was used for the inputs and weights and each neuron was pipelined after every stage of multiplication and addition. The average power consumption per neuron was ~1.06mW.

To summarize, the letter proposes a SOT-based device model for the design of an energy efficient artificial neuron. System level simulations of the electronic neuron with programmable resistive

1010

synapses suggest a possibility of ~3X improvement in power consumption compared to 45 nm CMOS technology.

**Acknowledgements:**

The work was supported in part by Center for Spintronic Materials, Interfaces, and Novel Architectures (C-SPIN), a MARCO and DARPA sponsored StarNet center, by Semiconductor Research Corporation, and by National Science Foundation.

**References:**

[1] Widrow, B. et al. "30 Years of Adaptive Neural Networks: Perceptron, Madaline, and Bacpropagation," *Proc. Of the IEEE* **78**, 1415-1442 (1990).

[2] George, D. et al. "The HTM Learning Algorithms," *Numenta tech. report* (2007).

[3] Cassidy, S. A. et al. "Design of silicon brains in the nano-CMOS era: Spiking neurons, learning synapses and neural architecture optimization," *Neural Networks* **45**, 4-26 (2013).

[4] Khan, M. et al. "SpiNNaker: mapping neural networks onto a massively-parallel chip multiprocessor," *In Proceedings of the IEEE International Joint Conference on Neural Networks (IJCNN),* 2008, pp. 2850–2857.

[5] Merolla, P. et al. "A Digital Neurosynaptic Core Using Embedded Crossbar Memory with 45pJ per Spike in 45nm," *In Proceedings of the IEEE Custom Integrated Circuits Conference (CICC),* 2011, pp. 1-4.

[6] Sharad, M. et al. "Proposal For Neuromorphic Hardware Using Spin Devices," *TECHCON 2012,* arXiv preprint arXiv:1206.3227 (2012).

[7] Sharad, M. et al. "Spin-Based Neuron Model with Domain Wall Magnets as Synapse," *IEEE Trans. On Nanotechnology* **11**, 843-853 (2012).

[8] Indiveri, G. et al. "Integration of nanoscale memristor synapses in neuromorphic computing architectures," *Nanotechnology* **24** 384010 (2013).

[9] Serrano-Gotarredona, T. et al. "A Proposal for Hybrid Memristor-CMOS Spiking Neuromorphic Learning Systems," *Circuits and Systems Magazine, IEEE* **13**, 74-88 (2013).

[10] Jo, S.H. et al. "Nanoscale Memristor Device as Synapse in Neuromorphic Systems," *Nano Letters* **10**, 1297-1301 (2010).

[11] Gao, B. et al. "Ultra-Low-Energy Three-Dimensional Oxide-Based Electronic Synapses for Implementation of Robust High-Accuracy Neuromorphic Computation Systems," *ACS Nano* **8,** 6998-7004 (2014).

[12] Slonczewski, J. C. "Conductance and exchange coupling of two ferromagnets separated by a tunneling barrier," *Phys. Rev. B* **39**, 6995–7002 (1989).

[13] Berger, L. "Emission of spin waves by a magnetic multilayer traversed by a current," *Phys. Rev. B* **54**, 9353 (1996)

[14] Hirsch, J. E. "Spin Hall effect," *Phys. Rev. Lett.* **83**, 1834–1837 (1999).

[15] Miron, I. M. et al. "Current-driven spin torque induced by the Rashba effect in a ferromagnetic metal layer," *Nature Mater.* **9**, 230–234 (2010).

[16] Miron, I. M. et al. "Perpendicular switching of a single ferromagnetic layer induced by in-plane current injection," *Nature* **476**, 189–193 (2011).

[17] Suzuki, T. et al. "Current-induced effective field in perpendicularly magnetized Ta/CoFeB/MgO wire," *Appl. Phys. Lett.* 98, 142505 (2011).

[18] Liu, L. et al. "Spin–torque switching with the giant spin Hall effect of tantalum," *Science* **336**, 555–558 (2012).

[19] Liu, L., Lee, O. J., Gudmundsen, T. J., Ralph, D. C. & Buhrman, R. A. "Current-induced switching of perpendicularly magnetized magnetic layers using spin torque from the spin Hall effect," *Phys. Rev. Lett.* **109**, 096602 (2012).

[20] Haazen, P. P. J. et al. "Domain wall depinning governed by the spin Hall effect," *Nature Mater.* **12**, 299–303 (2013).

[21] Kim, J. et al. "Layer thickness dependence of the current induced effective field vector in Ta/CoFeB/MgO," *Nature Mater*. 12, 240–245 (2013).

[22] Imre, A. et al. "Majority logic gate for magnetic quantum-dot cellular automata," *Science* **311**, 205–208 (2006).

[23] Salahuddin, S. and Datta, S. "Interacting systems for self-correcting low power switching," *Appl. Phys. Lett.* **90**, 093503 (2007).

[24] Lambson, B., Carlton, D. & Bokor, J. "Exploring the thermodynamic limits of computation in intergrated systems: magnetic memory, nanomagnetic logic, and the Landauer limit," *Phys. Rev. Lett.* **107**, 010604 (2011).

[25] Bhowmik, D et al. "Spin Hall effect clocking of nanomagnetic logic without a magnetic field," *Nature Nanotech.* **9**, 59 (2014)

[26] Behin-Aein, B., Datta, D., Salahuddin, S. and Datta, S. "Proposal for an all-spin logic device with built-in memory," *Nature Nanotech.* **5**, 266–270 (2010).

[27] Behin-Aein, B. et al., "Switching energy delay of all spin logic devices," *Appl. Phys. Lett.*, **98**, 123510 (2011).

[28] Beleggia, M. et al., "Demagnetization factors for elliptic cylinders," *J. Phys. D: Appl. Phys.*, **38**, 3333 (2005).

[29] Brink, A. et al., "Spin-Hall-assisted magnetic random access memory," *Appl. Phys. Lett.*, **104**, 012403 (2014).

[30] Fong, X. et al., "KNACK: a hybrid spin-charge mixed-mode simulator for evaluating different genres of spin-transfer torque MRAM bit-cells," *in Proc. of Int. Conf. on Simulation of Semicond. Processes and Dev. (SISPAD)*, 2011, pp. 51-54.

[31] LeCun, Y. et al., "Gradient-based learning applied to document recognition," *Proceedings of the IEEE*, **86,** 2278-2324 (1998).